\documentclass[journal,twoside,web]{IEEEtran}

\usepackage{cite}
\usepackage{amsmath,amssymb,amsfonts}
\usepackage{subcaption}
\usepackage{algorithmic}
\usepackage{graphicx}
\usepackage{textcomp}
\usepackage{multicol}
\usepackage{multirow}
\usepackage{array}
\usepackage{amsmath}
\usepackage{enumitem}

\def\BibTeX{{\rm B\kern-.05em{\sc i\kern-.025em b}\kern-.08em
    T\kern-.1667em\lower.7ex\hbox{E}\kern-.125emX}}
\begin{document}
\title{MAUS: A Dataset for Mental Workload Assessment on N-back Task Using Wearable Sensor}
\author{Win-Ken Beh, \IEEEmembership{Member, IEEE}, Yi-Hsuan Wu, and An-Yeu (Andy) Wu, \IEEEmembership{Fellow, IEEE}
\thanks{This work was supported in part by the Ministry of Science and Technology of Taiwan under grant MOST-109-2622-8-002-012-TA, and in part by PixArt Imaging Inc., Hsinchu, Taiwan, under grant Pix-108053. }
\thanks{Win-Ken Beh and Yi-Hsuan Wu are with 
the Electrical Engineering and the Graduate Institute of Electronics Engineering, NTU (e-mail: kane@access.ee.ntu.edu.tw; rick@access.ee.ntu.edu.tw).}
\thanks{An-Yeu (Andy) Wu is with 
the Electrical Engineering and the Graduate Institute of Electronics Engineering, NTU (e-mail: andywu@ntu.edu.tw).}

}

\maketitle
\begin{abstract}
This paper describes an open-access database focusing on the study of mental workload (MW) assessment system for wearable devices. A wristband photoplethysmogram (PPG) was provided as a representative of wearable devices. In addition, a clinical device that can
record Electrocardiography (ECG) , galvanic skin response (GSR) and, fingertip PPG was included in the database as a reference. The MW was induced by performing the N-back task with 22 subjects. The participants were asked to answer the Pittsburgh Sleep Quality Index (PSQI) questionnaire at the beginning of the experiment and the NASA Task Load Index (NASA-TLX) questionnaire after each N-back task. The result of data analysis show the potential uses of the recorded modalities and the feasibility of the MW elicitation protocol. Finally the MAUS dataset is now available for academic use\footnote[1]{The MAUS dataset is available at IEEE Dataport: https://ieee-dataport.org/open-access/maus-dataset-mental-workload-assessment-n-back-task-using-wearable-sensor}. Besides, we also presents a reproducible baseline system as a preliminary benchmark\footnote[2]{The code of the baseline system on MAUS dataset is available on Github: https://github.com/rickwu11/MAUS\_dataset\_baseline\_system}, which testing accuracy are 71.6 \%, 66.7 \%, and 59.9 \% in ECG, fingertip PPG, wristband PPG, respectively.

\end{abstract}

\begin{IEEEkeywords}
Electrocardiogram (ECG), photoplethysmogram (PPG), galvanic skin response (GSR), wearable sensors, mental workload, n-back, open access dataset
\end{IEEEkeywords}

\section{Introduction}
\label{sec:introduction}

Mental Workload (MW) refers to the portion of operator information processing capacity or resources that are required to meet system demands \cite{eggemeier1991workload}. A high mental workload means a more significant amount of information processing capacity or resources in performing a task. The MW assessment helps understand human operators in terms of processing capability or subjective psychological experiences \cite{lysaght1989operator}. Hence, MW assessment is an essential consideration for avoiding task error or working under "overload" conditions. It possesses numerous applications, ranging from safety to smart technology, including driver awareness, mental health monitoring, and Brain-Computer Interfacing (BCI). 

Mental workload is normally assessed through several physiological signals, such as electroencephalogram (EEG)\cite{liu2019entropy,chakladar2020eeg,pei2020eeg,fan2017eeg,lim2018stew}, electrocardiogram (ECG)\cite{cinaz2013monitoring}, galvanic skin response (GSR)\cite{schaule2018employing}, photoplethysmogram (PPG)\cite{jaiswal2019unobtrusive,ekiz2019long}. EEG-based methods offer an accurate assessment of MW\cite{hogervorst2014combining} with complex settings, while ECG, PPG, and GSR-based methods offer a moderate accuracy with simpler settings. Hence, we believe that ECG, PPG, and GSR-based methods will be the main solution in applying continuous MW assessment.   

Several works\cite{cinaz2013monitoring,schaule2018employing,jaiswal2019unobtrusive,ekiz2019long} have employed wearable sensors in assessing MW, but only validated on their in-house experimental database, usually without releasing data online. Hence, it is difficult for other researchers to make a fair comparison between methods. Thus, it motivated us to establish open access database for MW assessment with wearable sensors.

Our database consists of several physiological signals, includes single-lead ECG, GSR, fingertip PPG, and wristband PPG. We collected them under different MW conditions, which stimuli through the short-term memory task called N-back. A larger $N$ will stimuli a higher MW condition, while a smaller $N$ will do the opposite. $N$ is the golden objective reference of MW. We used it to discriminate different MW statuses. Besides that, for each subject, we had also collected the NASA Task Load Index questionnaire\cite{NASA-TLX} as a golden subjective reference of MW. 

One of the distinguishing features of our database is that we have collected synchronized wristband PPG, fingertip PPG, and ECG. To achieve long-term continuous monitoring of MW, wristband PPG is preferred. However, wristband PPG typically has worse signal quality than fingertip PPG. The proposed database provides both signals for a future feasibility study or wristband PPG enhancement for MW assessment applications.

In the following, we summarize the main contribution of the proposed dataset:
\begin{enumerate}
    \item We provide an open dataset consisting of the physiological signal acquired from wearable devices, allowing researchers to compare and validate their MW assessment method fairly.
    \item Our dataset provides synchronized ECG, fingertip-based PPG, and wrist-based PPG. Our dataset is the only dataset that provides synchronized ECG, wrist, and fingertips PPG to the best of our knowledge. Our dataset allows more comparative research between wrist and fingertip-based PPG in the application of MW assessment.
    \item We have provided a baseline system for MW assessment for future comparison (with code available). Thus, researchers can make fair comparisons based on our baseline system. 
\end{enumerate}

The remaining parts of this paper are organized as follows: Introduction and comparison between related databases will be presented in Section \ref{sec:related_databse}. Next, we introduce the experimental setup of the proposed database in Section \ref{sec:experiment_setup}. In Section \ref{sec:analysis}, we analyzed collected data and participant's subjective ratings to verify the correctness of the experiment flow. The baseline system and corresponding result will be introduced in Section \ref{sec:baseline}. Lastly, we will make some discussion on Section \ref{sec:discussion} and conclude this paper in Section \ref{sec:conclusion}.


\section{Related Database}
\label{sec:related_databse}

\begin{table*}[tbp]
\centering
\caption{Publicly available databases from related work}
\begin{tabular}{ccllc} 
\hline\hline
Database                    & Participants & \multicolumn{1}{c}{Stimuli} & \multicolumn{1}{c}{Collected Signals} & Signal Length/minutes  \\ 
\hline
CLAS\cite{CLAS}             & 62           & Math \& Logic Problem & ECG, GSR, PPG (Ear Lobe)               & 30                     \\
WAUC\cite{WAUC}             & 48           & MATB-II                     & EEG, ECG, GSR, PPG (Wrist), TEMP, ACC~ & 60                     \\
MMOD-COG\cite{MMOD}         & 40           & Arithmetic Task             & ECG,GSR                               & 14.3                   \\
Cognitive \& Physical Load\cite{physionet} & 13           & Auditive 2-back Task        & ECG, PPG (Fingertip), ACC              & 32.9  \\
CogLoad\cite{dataset_wearable}   & 23           & N-back Task                 & GSR,TEMP, ACC                         & 18                     \\
\textbf{Proposed Database}           & \textbf{22}           & \textbf{N-back Task}                 & \textbf{ECG, GSR, PPG (Fingertip), PPG (Wrist)}  & \textbf{35}                     \\
\hline\hline
\end{tabular}
\label{table:related_database}
\end{table*}

Our goal is to establish a database that allows researchers to develop their algorithms and compare results. Since we only focused on long-term MW monitoring scenarios. Hence, we considered only those easy-acquired physiological signals. In this case, we do not consider EEG-related databases, yet we compared the MW database that consists of some easy-acquired signals such that ECG, PPG, and GSR. To the best of our knowledge, few open-access databases collect these easy-acquired signals, which are,
\begin{itemize}
    \item CLAS \cite{CLAS}
    \item WAUC \cite{WAUC}
    \item MMOD-COG \cite{MMOD}
    \item Simultaneous physiological measurements with five devices at different cognitive and physical loads \cite{physionet}
    \item Datasets for Cognitive Load Inference Using Wearable Sensors and Psychological Traits \cite{dataset_wearable}
\end{itemize}
In this section, we provide a brief review of the literature related to the proposed database. Then, we will compare these databases based on stimuli selection, physiological modalities and, MW level annotations.

\subsection{Stimuli}

Databases \cite{CLAS,WAUC,MMOD,physionet,dataset_wearable} have different strategies in generating different MW conditions. Different stimuli can make huge differences between databases. Some stimuli can make obvious physiological response while other making no physiological changes. Hence, stimuli selection could be an essential factor in the practical usage of MW assessment. It also affects a lot in assessment performance.

CLAS \cite{CLAS} used a series of interactive tasks to elicit different MW levels. Those tasks include solving sequences of \textit{Math problems}, \textit{Logic problems}, and the \textit{Stroop test}. Participants in CLAS requested to answer those assignments in a short time. It claimed that these tasks are designed to evaluate different aspects of the momentary cognitive load.
Besides that, WAUC \cite{WAUC} employed MATB-II\cite{MATB-II} task to elicit two levels of MW. This set of tasks was initially devised to simulate different activities that need to be performed by an aircraft pilot.
On the other hand, MW in MMOD-COG\cite{MMOD} was induced by way of performing the primary arithmetic task. Participants in MMOD-COG were asked to perform two tasks with different complexity. The high complexity task presents 4-operand arithmetic operations, while the low complexity task uses 2-operand arithmetic operations.
Next, \cite{physionet} used the auditive 2-back task to elicit high MW status. By comparing 2-back and resting status, two different MW levels are obtained.
Furthermore, \cite{dataset_wearable} used the N-back task during the experiment. 2-back and 3-back tasks were used during the experiment session. The larger number of $N$ will elicit a higher MW level.

\subsection{Physiological Modalities}

Most MW databases provide EEG measurements for exploration. In this paper, we focused on easy-acquired physiological signals, which can be measure by smartwatches. The physiological measurements appeared in \cite{CLAS,WAUC,MMOD,physionet,dataset_wearable} are,
\begin{multicols}{2}
    \begin{itemize}
        \item EEG
        \item Skin Temperature
        \item GSR
        \item Fingertip PPG
        \item 3-Axis Acc.
        \item ECG
        \item Breathing Rate
        \item Wrist PPG
    \end{itemize}
\end{multicols}

These signals were monitored using wearable or off-the-shelf devices. These databases used different instruments for recording signals. Hence, the sampling rate varies among databases. On the other hand, different MW stimuli and experiment settings made the signal length in the database different. For example, each person in the database CLAS is represented with a 30-minutes recording of the physiological signal. In contrast, each person in WAUC represented a 60-minutes of the physiological signal. 

\subsection{Reference of Mental Workload Level}

Reference of MW level generally consists of subjective and objective assessment. A subjective MW assessment relies on participants periodically filling in a questionnaire with ratings related to their mental status. Popular questionnaires include the Subjective Workload Assessment Technique (SWAT), NASA Task Load Index (NASA-TLX), and Modified Cooper-Harper Scale. Then, we can obtain a quantitative score based on these questionnaires.

Objective MW reference regards stimuli tags during experiment steps. For example, the WAUC database used MATB-II as their MW stimuli. The modulation of MW relies on changing parameters in MATB-II. For the lower MW case, sliding bar speed, aim speed, the volume of fuel in the reservoirs, and failure rate of the pumps were set to lower values. These changes could be a stimuli tag that can be treated as an objective MW reference. Moreover, MMOD-COG used the stimuli tags of 2-operand and 4-operand arithmetic tasks as the reference low and high MW level, respectively.

At the end of this section, we summarized the related database in Table~\ref{table:related_database}. We have compared related databases to ours. The number of participants, stimuli selection, collected signal, and signal length of each participant are compared. Compared to other databases, our database provides synchronized wrist-based PPG, fingertip-based PPG, and ECG signals. These signals allow the researcher to make comparisons and study how to make wrist-based PPG better to achieve fingertip-based PPG or ECG performance. We think it will pave the way toward long-term MW monitoring.

\section{Experiment Setup}
\label{sec:experiment_setup}

In this section, we first introduce information about the participants performing this experiment. Then, the apparatus used for data acquisition is presented. Finally, the experimental procedures, including the self-assessment, are described in detail.
\subsection{Participants}
Our database contains 22 healthy participants (2 females)  from the university's graduate students. The average age among the participants was 23 years with a 1.7 standard deviation. Prior to the experiment, each subject signed a consent form. Next, they were informed of a set of instructions explaining experimental procedures and the meaning of the self-assessment questionnaires.
\subsection{Data Acquisition}
Participants were seated in a room with moderate temperature, controlled illumination, and noise, at a table with physiological signal recording equipment, a computer monitor for delivering tasks, and a keyboard for reacting to the stimuli.

\begin{figure}[tbp]
\centering
\includegraphics[width=\columnwidth]{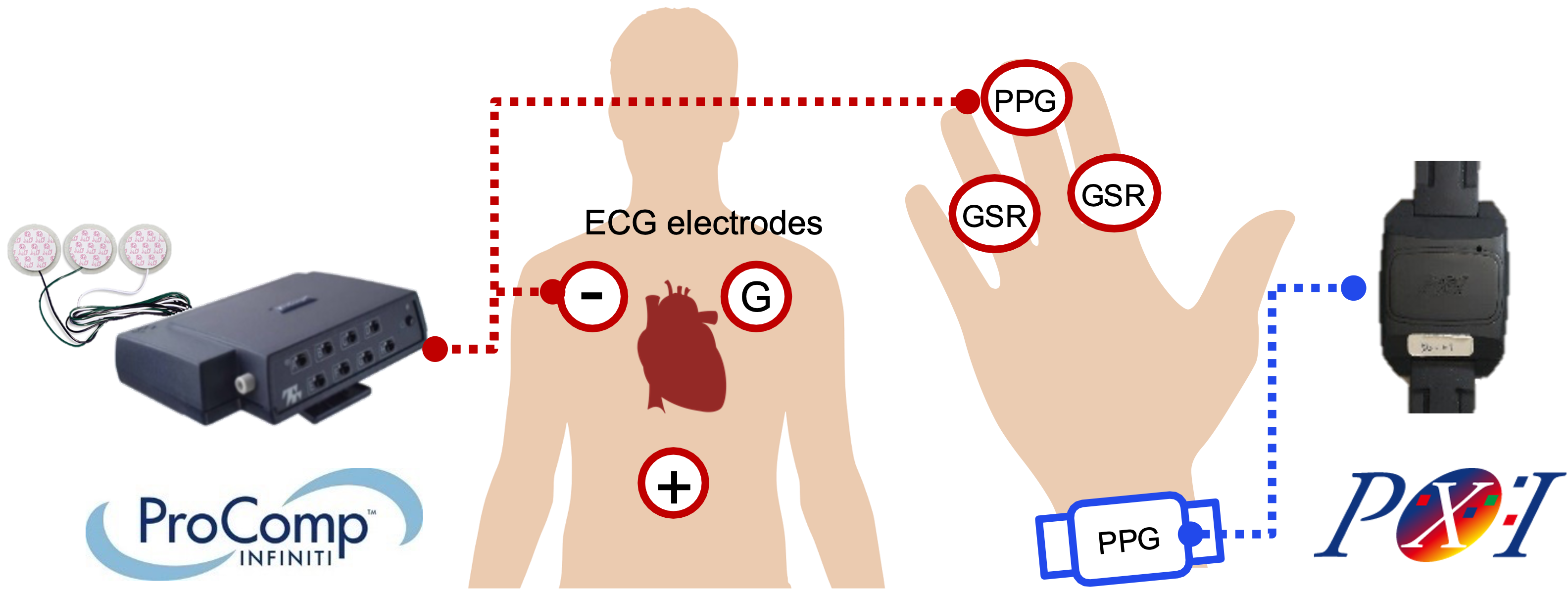}
\caption{Placement of peripheral physiological sensors. ECG electrodes placement GSR, fingertip PPG, and wrist-worn PPG placement (right). }
\label{fig:placement}
\end{figure}

In this database, two devices were employed to acquire four channels of data simultaneously, as summarized in Table~\ref{table:database}. The Procomp Infiniti can provide high signal fidelity for multi-channel measurement in real-time. It offers internal, user-activated calibration to ensure that the highest quality signal can be obtained, which can be viewed as a golden reference. The PixArt PPG watch provides the signal measured from a wearable device. It offers the opportunity to study data processing methods for using wearable devices. Fig.~\ref{fig:placement} illustrates the electrode placement for the acquisition of peripheral physiological signals. 

\subsubsection{Procomp Infiniti}
The Procomp Infiniti was used to record ECG, GSR, and PPG at a sampling rate of 256 Hz. Single lead ECG was measured by electrodes placed on both shoulders and one on the abdomen. GSR was measured via two electrode straps fastened around the index and ring finger of participants' non-dominant hands. PPG was measured by a sensor pressed against the palmar surface of the middle fingertip with an elastic strap fastened tightly to acquire better signal quality. 
\subsubsection{PixArt PPG Watch}
The PPG was recorded from wrist-type PPG sensors with green LED at a sampling rate of 100 Hz. The watch was worn on the non-dominant hand to avoid the movement of answering on the keyboard. The data would then be transmitted to a tablet through Bluetooth.

\begin{table}[tbp]
\centering
\caption{Database content summary}
\begin{tabular}{lll} 
\hline\hline
\multicolumn{3}{c}{\textbf{Participants and modalities}}                                                                                                       \\ 
\hline
\textbf{No. of participants}   & \multicolumn{2}{l}{22 (20 males, 2 females)}                                                                                                       \\
\textbf{Recorded signals}      & ProComp Infiniti & ECG (256 Hz)                                                                                    \\
                               &                                                & GSR (256 Hz)                                                                     \\
                               &                                                & PPG (256 Hz)                                                                     \\
                               & PixArt Watch                                  & PPG (100 Hz)                                                                     \\
\hline
\multicolumn{3}{c}{\textbf{Experiment setup}}                                                                                                       \\
\hline
\textbf{\textbf{N-back task}} & \multicolumn{2}{l}{0-back: low MW; 2,3-back: high MW}                                                                                                       \\
\textbf{\textbf{No. of trials}} & \multicolumn{2}{l}{6 (0, 2, 3, 2, 3, 0 -back)}                                                                                                       \\
\textbf{Trial duration}         & \multicolumn{2}{l}{7 min. (5 min. N-back + 2 min. Resting)}                                                                                                       \\
\textbf{Total duration}         & \multicolumn{2}{l}{47 min. (5 min. Resting + 7 min. * 6 trials)}                                                                                                       \\
\hline
\multicolumn{3}{c}{\textbf{Sleep evaluation (PSQI)}}                                                                                                       \\
\hline
\textbf{Rating scales}         & \multicolumn{2}{l}{\begin{tabular}[c]{@{}l@{}}Sleep quality, Sleep latency, Sleep duration, \\Habitual sleep efficiency, Sleep disturbances, \\Sleeping medication, Daytime dysfunction\end{tabular}}  \\
\textbf{Rating values}         & \multicolumn{2}{l}{0-21 (each 0-3)}                                                                                                       \\
\hline
\multicolumn{3}{c}{\textbf{Subjective MW rating (NASA-TLX)}}                                                                                                       \\
\hline
\textbf{Rating scales}         & \multicolumn{2}{l}{\begin{tabular}[c]{@{}l@{}}Mental demand, Physical demand, Temporal\\demand,~Performance, Effort, Frustration\end{tabular}}  \\
\textbf{Rating values}         & \multicolumn{2}{l}{0-100 with step of 5}                                                                                                       \\
\hline\hline
\end{tabular}
\label{table:database}
\end{table}

\begin{figure*}[tbp]

 \center

  \includegraphics[width=\textwidth]{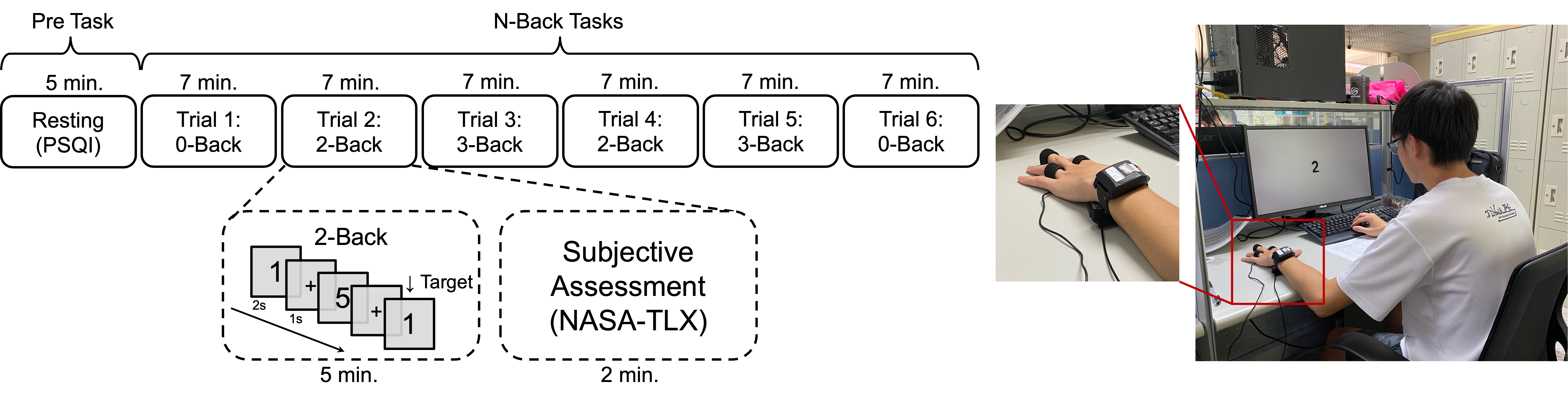}

  \caption{Experiment procedure and subject setup.}

  \label{fig:experiment flow}

\end{figure*}

\subsection{Experimental Procedure}
The entire experiment procedure with session duration is summarized in Fig.~\ref{fig:experiment flow}. It contains three kinds of sessions:

\subsubsection{Pre-task Pittsburgh Sleep Quality Index (PSQI)}
At the start of the experiment, there would be five minutes resting session, and participants were asked to fill out the Pittsburgh Sleep Quality Index (PSQI)~\cite{PSQI} questionnaires, commonly used to assess a subject's sleep quality over the past one month. The PSQI contains seven subscales, including subjective sleep quality, sleep latency, sleep duration, habitual sleep efficiency, sleep disturbances, use of sleeping medication, and daytime dysfunction. The component scores range from 0 to 3 and are summed to give a total score ranging from 0 to 21. The higher PSQI score suggests more significant sleep disturbance, and a global score higher than 5 indicates a significant disturbance. Previous studies have shown the validity and reliability of the Persian version PSQI\cite{moghaddam2012reliability}.
\subsubsection{N-back Task}
After the resting session, there would be six trials of the N-back task~\cite{n-back}. In the N-back task, the user had to memorize the last $N$ one-digit numbers of a series of rapidly flashing numbers in succession. When a stimulus was identical to the \emph{n}-th number preceding the stimulus number, the subject had to respond by pressing the space bar on the keyboard. As $N$ increased, the task's difficulty increased, as subjects had to memorize more numbers and keep shifting the memorized sequence. Therefore, adjusting the $N$ could stimulate different levels of MW. There would be 0-, 2-, and 3-back three kinds of tasks. The 0-back task could be viewed as a lower MW state, while the 2- and 3- back tasks were regarded as a higher MW state. These three kinds of tasks were arranged in a counterbalanced order (i.e., 0→2→3→2→3→0)~\cite{shin2018simultaneous} for a total of six trials. A single trial began with instruction showing the type of task(0-, 2- or 3-back). After the subject pressed the space bar for continuing, the trial presented a random one-digit number every 3 s. Each number was displayed for 2 s, followed by a fixation cross for the remaining 1 s. A total of 100 numbers were presented in a trial with 25 targets (25\%). The answers and the response time were recorded for performance evaluation. Each task was followed by two minutes resting period for the subjective rating of the MW of the task.
 
\subsubsection{Post-task NASA-TLX}
As mentioned above, each n-back task was followed by a subjective evaluation. In this step, NASA Task Load Index (NASA-TLX) questionnaire~\cite{NASA-TLX} was adopted. The NASA-TLX is a widely used, multidimensional subjective assessment tool for rating the perceived load of doing a task. The NASA-TLX questionnaire consists of six subscales to evaluate MW: mental demand, physical demand, temporal demand, performance, effort, and frustration. Each subscale is evaluated from 0 to 100 in intervals of 5. After the six scores were obtained, the subject had to compare the six subscales pairwise based on their perceived importance to create individual weighting on those subscales. The number of times each subscale was selected was its weight. Each subscale score was multiplied by the corresponding weight, summed, and then divided by 15 to obtain a workload score from 0 to 100 as the global score. The score represented the overall subjective rating of MW in the task.
\section{Data Analysis}
\label{sec:analysis}
This section provides the analysis of the collected data, including analysis of the pre-task questionnaire, user performance, subjective rating, and psychophysiological recordings.

\begin{figure*}[tbp]
    \centering
    \subfloat[]{\label{sublable1}\includegraphics[scale=0.15]{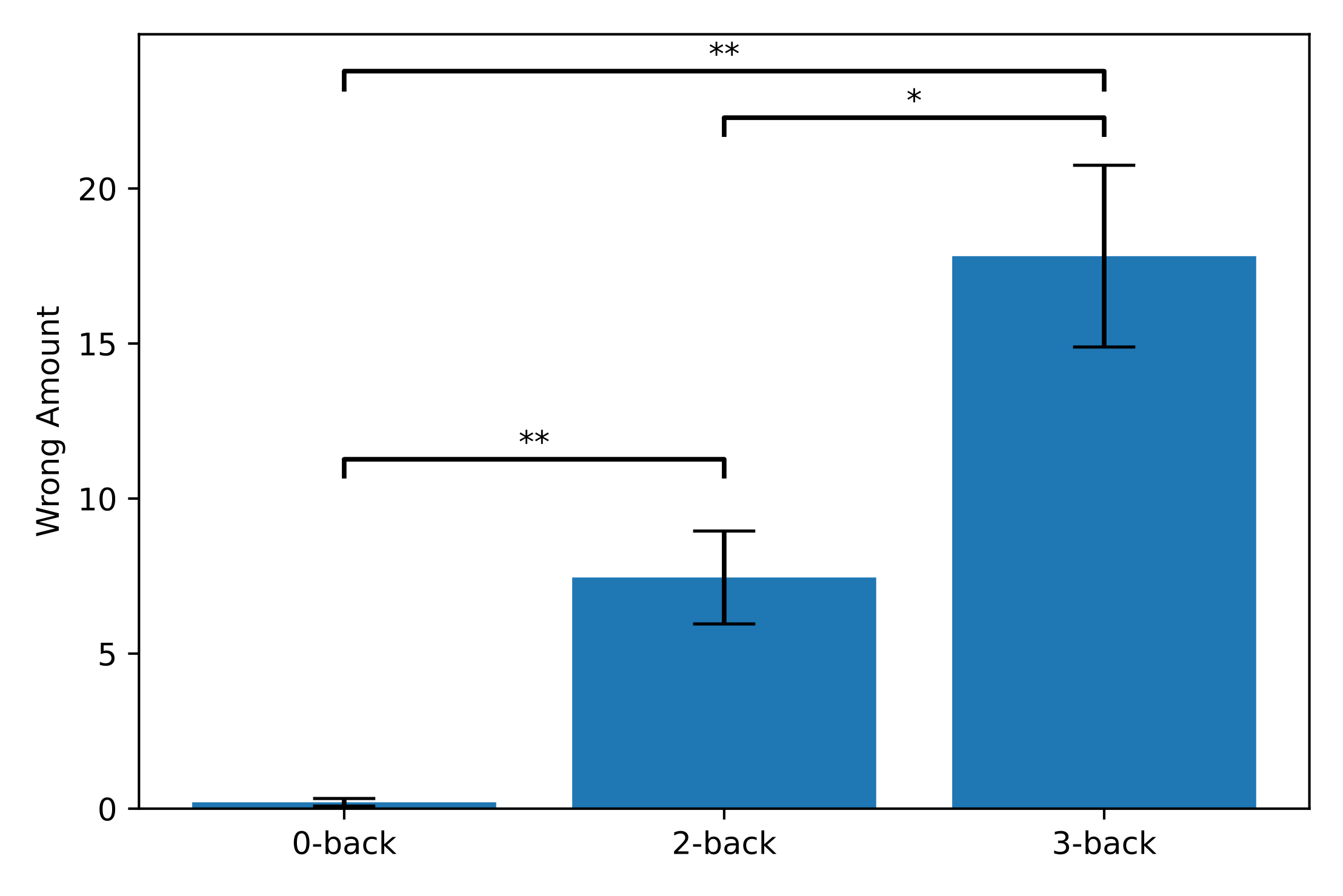}} 
    \subfloat[]{\label{sublable2}\includegraphics[scale=0.15]{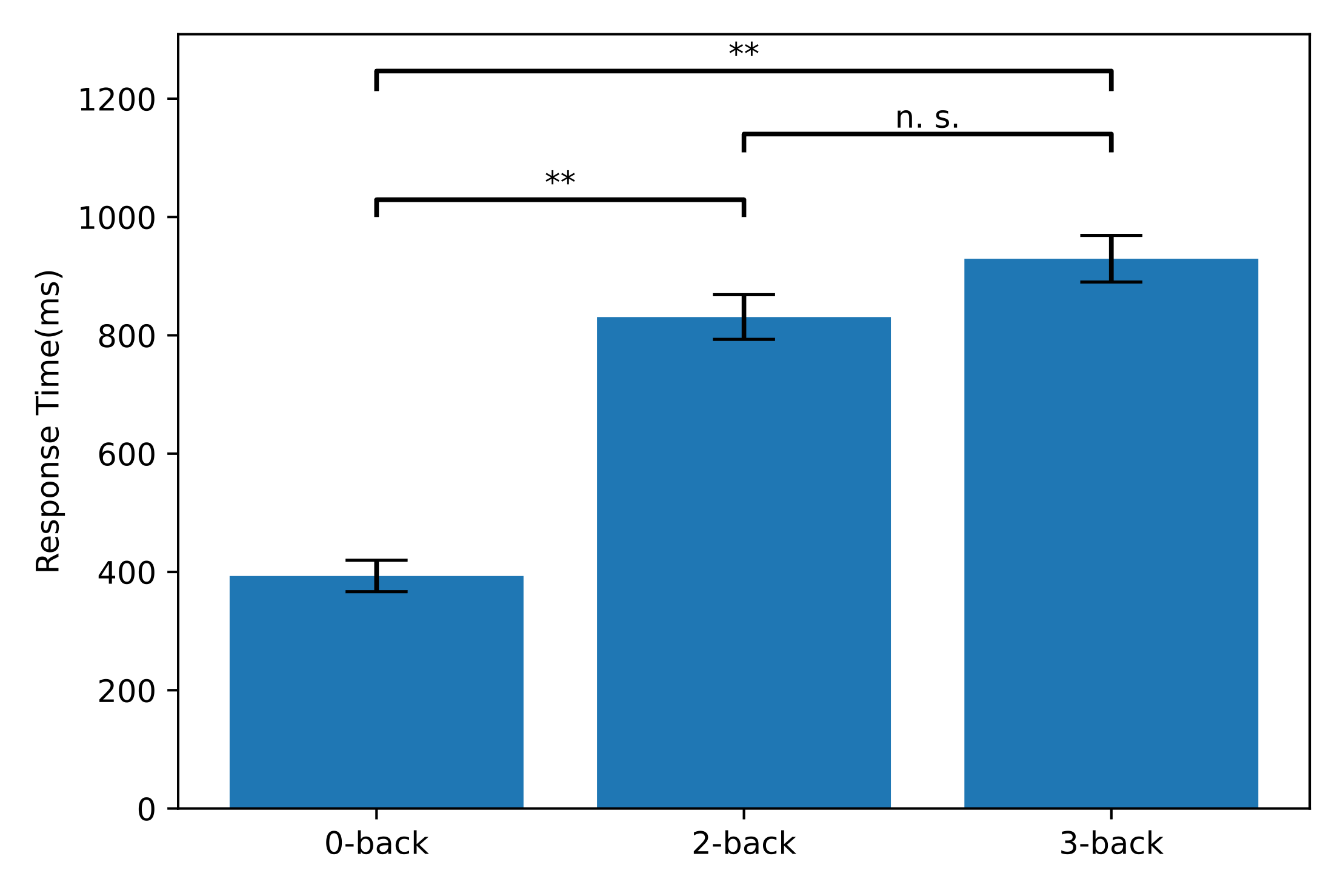}} 
    \subfloat[]{\label{sublable2}\includegraphics[scale=0.15]{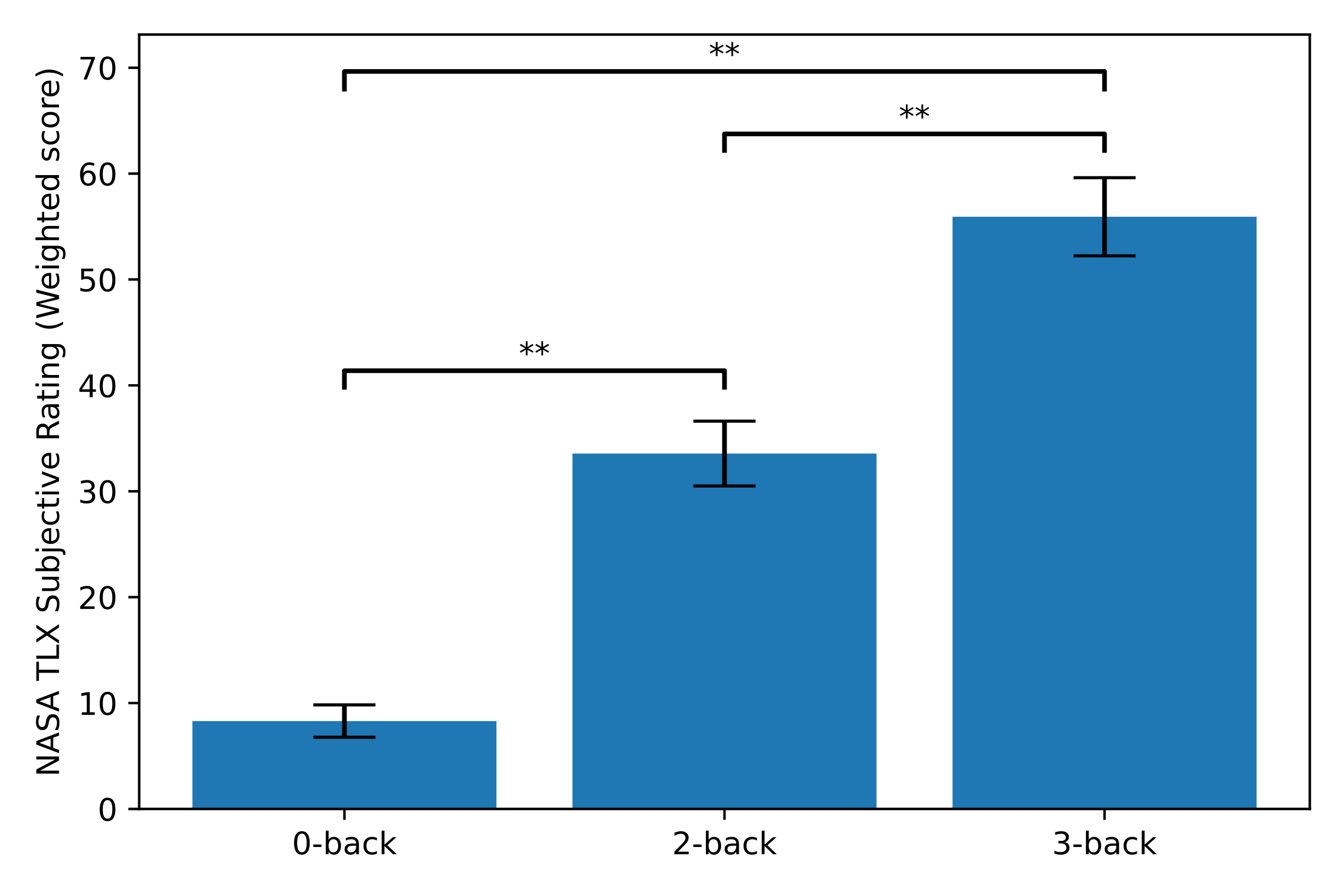}} 
    \caption{User performance and subjective evaluation in the n-back tasks (a) Average wrong amount, (b) Average response time, (c) Average subjective evaluation of task difficulty (NASA-TLX score). Significance is represented by **: \emph{p}-value $<$ 0.001, *: 0.001 $\leq$ \emph{p}-value $<$ 0.05, n.s.: \emph{p}-value $\geq$ 0.05. Error bars indicate standard errors of the mean.}
    \label{fig:performance}
\end{figure*}

\begin{figure*}[tbp]
    \centering
    \subfloat[]{\label{sublable1}\includegraphics[scale=0.16]{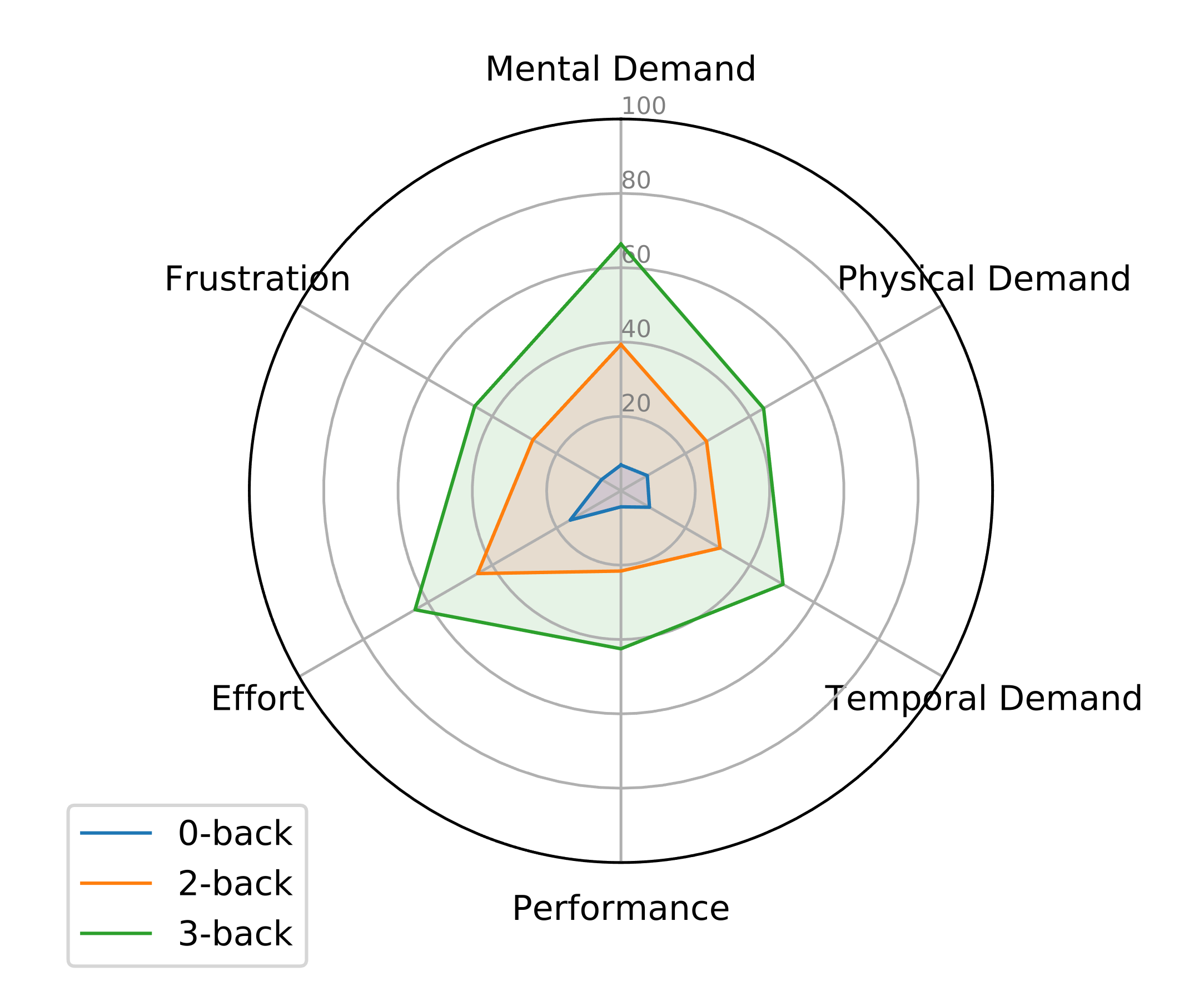}} 
    \subfloat[]{\label{sublable2}\includegraphics[scale=0.16]{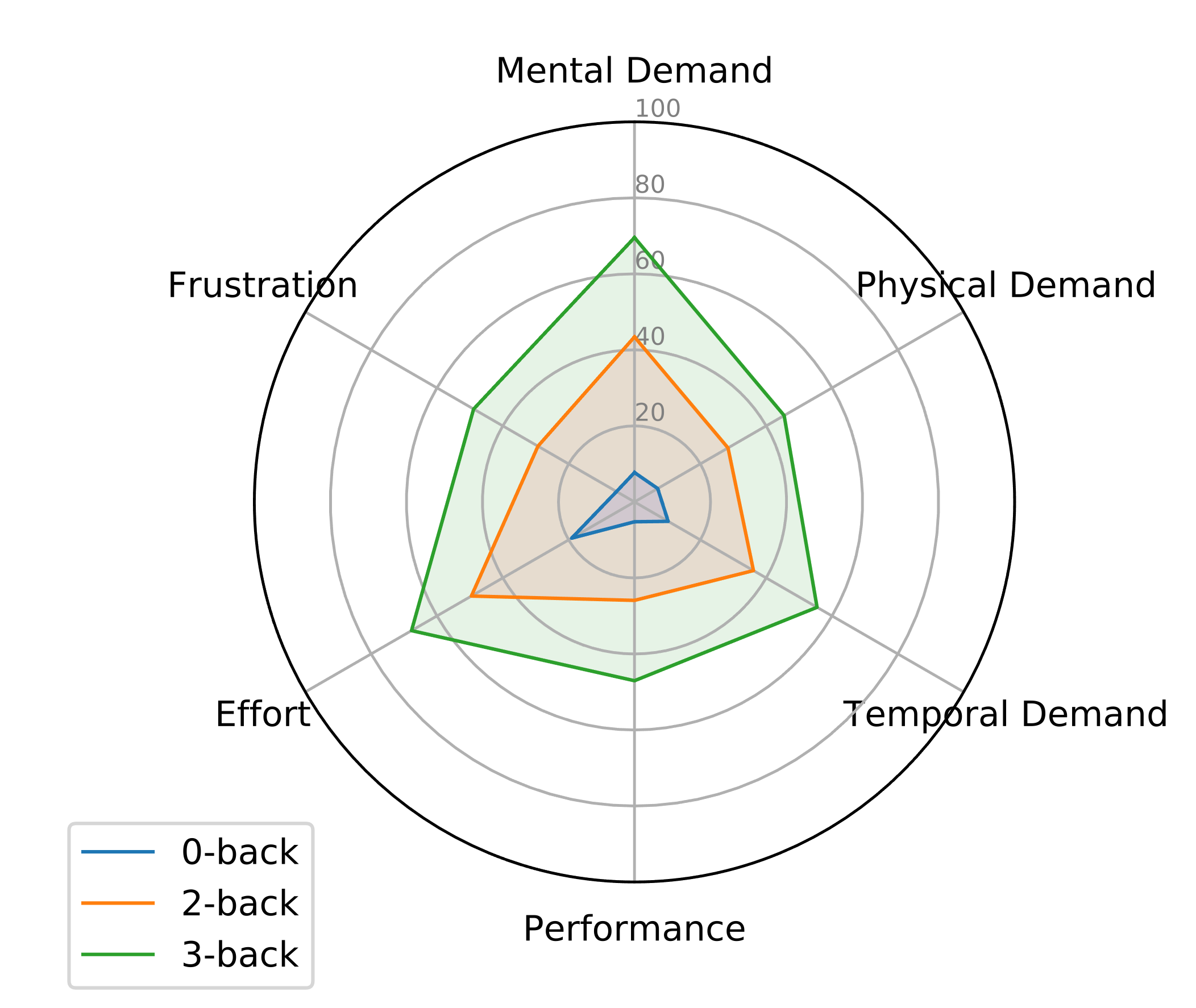}} 
    \subfloat[]{\label{sublable2}\includegraphics[scale=0.16]{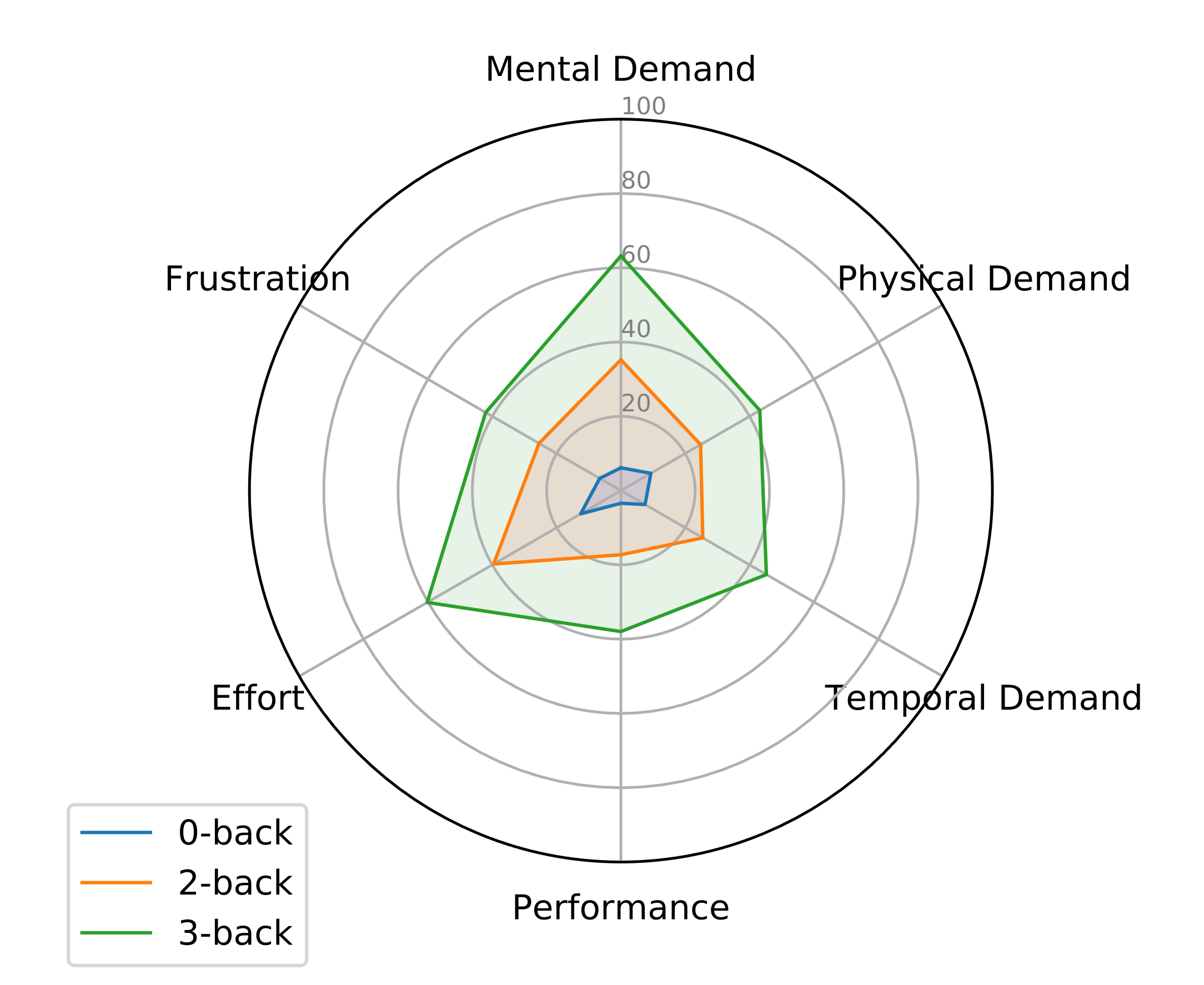}} 
    \caption{Distribution of NASA TLX Dimensions of (a) Mean of two round (0-,2-,3-,2-,3-,0-) , (b) First round of (0-,2-,3-) trials, (c) Second round of (2-,3-,0-) trials.}
    \label{fig:NASATLX}
\end{figure*}

\subsection{User Performance and Subjective Rating Analysis}
To verify that subjects perceived different mental workload states through different n-back conditions, we analyzed user performance and subjective ratings.
Fig.~\ref{fig:performance} shows user performance and subjective evaluation of the experiment, including mean values and standard deviation. 

\subsubsection{Performance measurement}
We evaluated the wrong amount as the performance in the n-back task. The one-way analysis of variance (ANOVA) indicated significant differences in the wrong amount for the three n-back levels (\emph{p} $<$ 0.05). The wrong amount included missed target and incorrect responses, respectively, a subject failed to press the key when a target stimulus was present, and the subject incorrectly identified a letter as a target. The wrong amount increased from a mean of 0.2 (times) for the 0-back to 7.5 for the 2-back to 17.8 for the 3-back task, which clearly indicates that the three tasks differ significantly in difficulty. We also evaluated the reaction time, i.e., when a subject pressed the space bar when the target appeared. The response time is significantly influenced by the n-back level. The difference between 0 and 2-back and 0 and 3-back are significant (ANOVA, \emph{p} $<$ 0.001), while the difference between 2 and 3-back is not statistically significant. These two performance-related factors suggest that n-back tasks with higher $N$ can stimulate higher MW states.

\subsubsection{Subjective measurement}
A weighted score of NASA-TLX was calculated from the six subscales, representing the overall workload of subjective rating. The ANOVA of NASA-TLX score showed a significant difference between the three levels of n-back tasks (\emph{p} $<$ 0.001). Post hoc comparisons indicated that MW during high n-back (2,3-back) tasks was significantly greater than those during the low n-back task (0-back), suggesting n-back tasks with higher $N$ require higher mental workload than those with lower $N$. In addition, the mean value of the six NASA-TLX subscales was showed in Fig.~\ref{fig:NASATLX}, considering the MW from low to high (0-,2-,3-back) and representing in blue, orange, and green, respectively. The Radar charts provided us with a visualization of the six-dimensional distributions in NASA-TLX. Overall, all six subscales grew together as $N$ increases. We can also found a slight degradation in scores between the first Fig.~\ref{fig:NASATLX}(b) and second-round Fig.~\ref{fig:NASATLX}(c).



\subsection{Pittsburgh Sleep Quality Index (PSQI) Analysis}
\begin{figure}[tbp]
\centering
\includegraphics[scale=0.18]{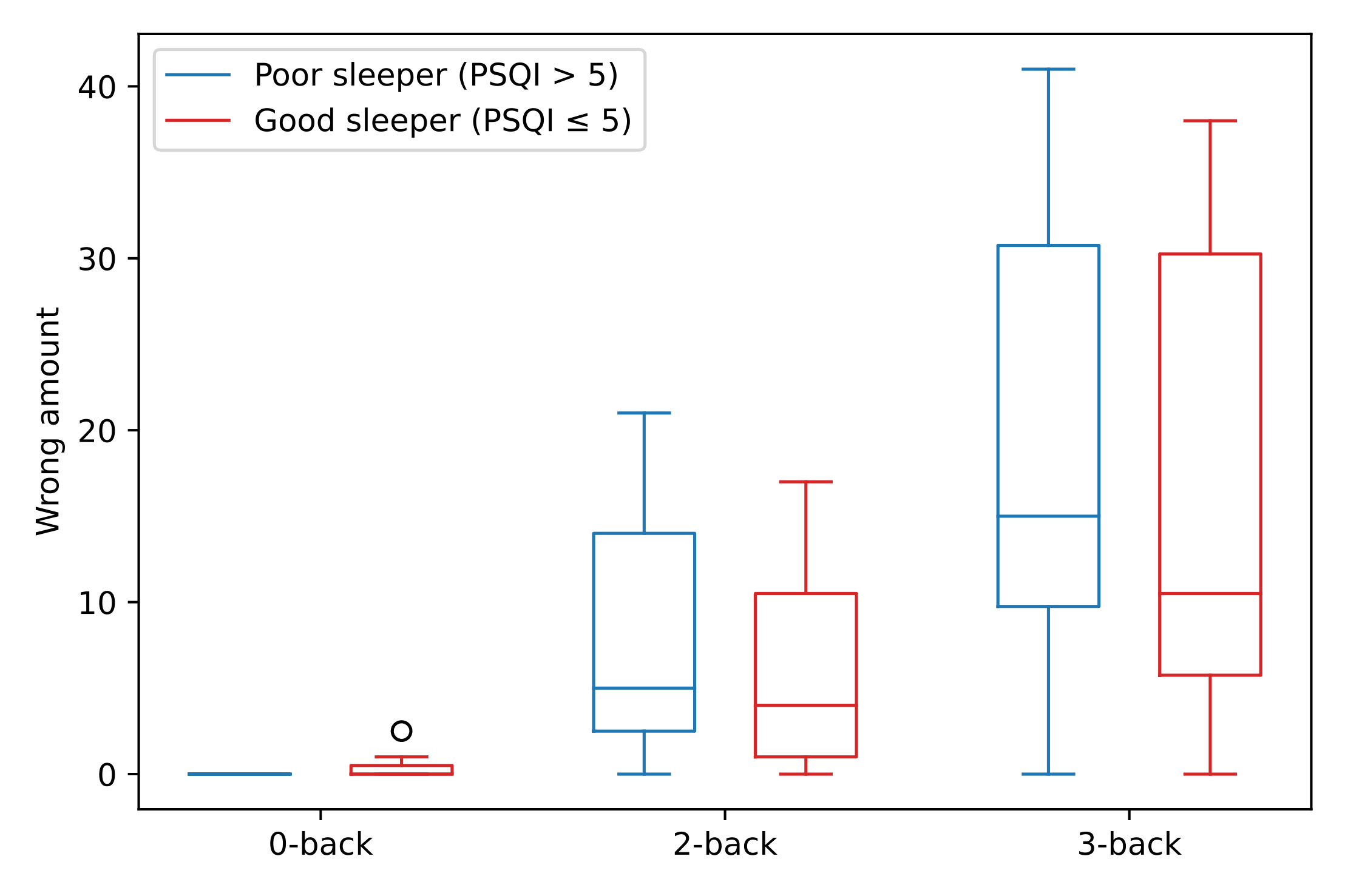}
\caption{N-back wrong amount between poor and good sleeper.}
\label{fig:psqi_performance}
\end{figure}

The subjective sleep quality scores of each subject were listed in Table~\ref{table:psqi}. Ten subjects were reported as having poor sleep quality based on the PSQI (PSQI $>$ 5). To investigate the relationship between subjective sleep quality and cognitive performance, we analyzed the wrong amount in the n-back task for the poor sleepers and good sleepers, as shown in Fig.~\ref{fig:psqi_performance}. Those poor sleepers tended to have more false reactions in the n-back tasks than those good sleepers. However, there was no  statistical significance between the two groups in both the 2-back task (ANOVA, \emph{p} = 0.46) and the 3-back task (ANOVA, \emph{p} = 0.68), indicating that the subjects' data in this database were not affected by sleep quality and were valid. Also, Pearson's r correlation coefficient shows no correlation between the wrong amount and the PSQI scores (r = 0.056 in 2-back task and r = 0.079 in 3-back task). 

\begin{table}[tbp]
\centering
\caption{Participants' PSQI}
\begin{tabular}{|c|c|c|c|}
\hline
\textbf{Subject ID} & \textbf{PSQI} & \textbf{Subject ID} & \textbf{PSQI} \\ \hline
002                 & 5             & 015                 & 5             \\
003                 & 9             & 016                 & 7             \\
004                 & 6             & 017                 & 6             \\
005                 & 6             & 018                 & 12            \\
006                 & 1             & 019                 & 7             \\
008                 & 5             & 020                 & 2             \\
010                 & 4             & 021                 & 6             \\
011                 & 7             & 022                 & 4             \\
012                 & 4             & 023                 & 5             \\
013                 & 11            & 024                 & 2             \\
014                 & 0             & 025                 & 2             \\ \hline
\end{tabular}
\label{table:psqi}
\end{table}

\subsection{Comparison between Wrist and Fingertip PPG}
\begin{figure}
\centering
\includegraphics[width=\columnwidth]{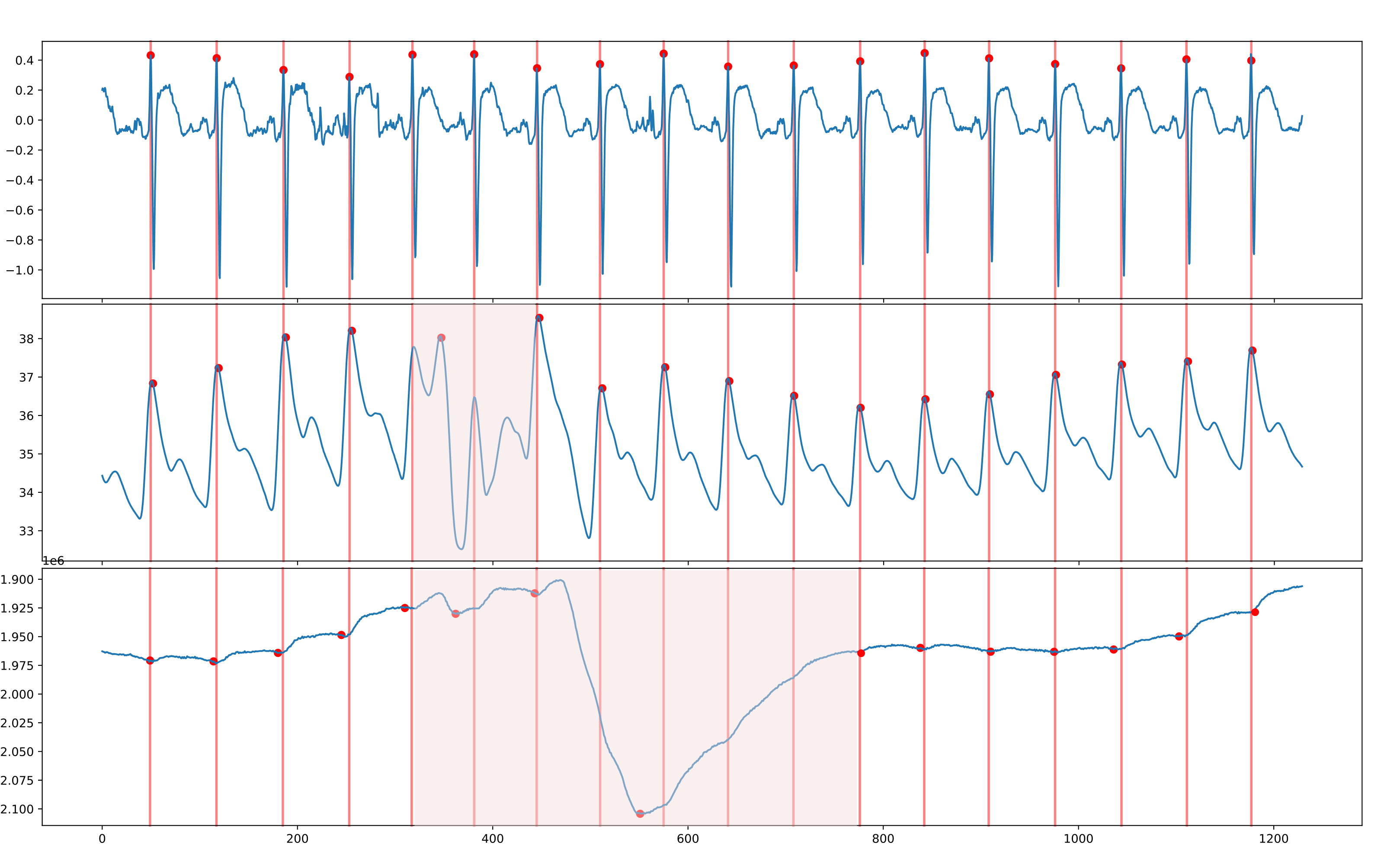}
\caption{Comparison between 3 channel signals with detected peaks/valleys, red translucent regions represent incorrect inter-beat intervals comparing to ECG: ECG (top), PPG (fingertip) (middle), PPG (wrist-worn) (bottom).}
\label{fig:comparison}
\end{figure}

In our dataset, ECG, fingertip PPG, and wristband PPG were collected simultaneously, providing the opportunity for comparative studies between these signals. PPG is more prone to motion artifacts and sensor positioning comparing to ECG. In addition, wristband PPG tends to have worse signal quality than fingertip PPG and suffers various distortions. Fig.~\ref{fig:comparison} shows three signals with about 12 seconds acquired during a five minutes n-back task. For the ECG, the inter-beat intervals (IBI) extracted from R-peaks could be considered as a golden reference since the electrode measurement was stable in this experiment. The IBIs from fingertip PPG were slightly affected by the motion, while IBIs from wristband PPG were more incorrect due to the baseline wandering caused by motion. The incorrect IBIs could lead to accuracy degradation of HRV analysis.

\subsection{Statistical Analysis on HRV Features}
The ANOVA was adopted for statistical analysis of all extracted HRV features. It was used to determine which HRV features could discriminate between low and high MW states (\emph{p} $<$ 0.05). We extracted 13 commonly used time and frequency domain features recommended in~\cite{camm1996heart}, details of which are presented in the next section. The \emph{p}-value for each feature between low and high MW (0 vs. 2-back, 0 vs. 3-back) was presented in Table~\ref{table:anova}, including three channel signals for comparison. In both scenarios (0 vs. 2-back and 0 vs. 3-back), significant differences could be observed for the features \emph{SDNN} and \emph{HF} of ECG and fingertip PPG, while no statistical differences were observed in these two features of wrist PPG. Overall, it could be observed that the statistical significance of HRV features for ECG and fingertip PPG were consistent. However, the statistical significance of HRV features from wrist PPG showed a more inconsistent trend than the other two signals, reflecting the signal quality difference of the three signals described in the previous subsection. That is, the signal of fingertip PPG was more stable than wrist PPG, which made the features extracted from fingertip PPG closer to ECG (golden reference). 

\begin{table*}[tbp]
\centering
\caption{Analysis of variance (ANOVA) for HRV features of the three modalities}
\resizebox{\linewidth}{!}{%
\begin{tabular}{>{\centering\hspace{0pt}}m{0.115\linewidth}>{\hspace{0pt}}m{0.104\linewidth}>{\centering\hspace{0pt}}m{0.115\linewidth}>{\centering\hspace{0pt}}m{0.119\linewidth}>{\centering\hspace{0pt}}m{0.119\linewidth}>{\centering\hspace{0pt}}m{0.119\linewidth}>{\centering\hspace{0pt}}m{0.119\linewidth}>{\centering\arraybackslash\hspace{0pt}}m{0.119\linewidth}} 
\hline\hline
                                                              & \multicolumn{1}{>{\centering\hspace{0pt}}m{0.104\linewidth}}{} & 0 vs 2 back                                &                                      &                             & 0 vs 3 back                                 &                                      &                              \\ 
\hline
                                                              & Feature                                                        & ECG                                  & Fingertip PPG                           & Wrist PPG                  & ECG                                  & Fingertip PPG                            & Wrist PPG                     \\ 
\hline
\multicolumn{1}{>{\hspace{0pt}}m{0.115\linewidth}}{Time}      & SDNN                                                           & \textbf{0.002760\rlap{\textsuperscript{*}}} & \textbf{0.000900\rlap{\textit{\textsuperscript{**}}}} & 0.636859                    & \textbf{0.000017\rlap{\textit{\textsuperscript{**}}}} & \textbf{0.000041\rlap{\textit{\textsuperscript{**}}}} & 0.593607                     \\
                                                              & NN50                                                           & 0.121902                             & 0.148574                             & 0.511712                    & \textbf{0.007757\rlap{\textsuperscript{*}}}          & 0.142155                             & 0.104222                     \\
                                                              & PNN50                                                          & 0.099391                             & 0.102797                             & 0.390081                    & \textbf{0.011236\rlap{\textsuperscript{*}}}          & 0.109180                             & \textbf{0.049078\rlap{\textsuperscript{*}}}  \\
                                                              & RMSSD                                                          & 0.062794                             & \textbf{0.031618\rlap{\textsuperscript{*}}}          & 0.790807                    & \textbf{0.010403\rlap{\textsuperscript{*}}}          & 0.225226                             & 0.552315                     \\
                                                              & SDSD                                                           & \textbf{0.049761\rlap{\textsuperscript{*}}}          & \textbf{0.023827\rlap{\textsuperscript{*}}}          & 0.870443                    & \textbf{0.003506\rlap{\textsuperscript{*}}}          & 0.411065                             & 0.290467                     \\
                                                              & TINN                                                           & 0.503672                             & 0.348718                             & 0.549263                    & \textbf{0.039293\rlap{\textsuperscript{*}}}          & \textbf{0.003889\rlap{\textsuperscript{*}}}          & 0.257971                     \\
                                                              & TRI Index                                                      & 0.080009                             & 0.083000                             & 0.211993                    & \textbf{0.011210\rlap{\textsuperscript{*}}}          & \textbf{0.005751\rlap{\textsuperscript{*}}}          & \textbf{0.001868\rlap{\textsuperscript{*}}}  \\ 
\hline
\multicolumn{1}{>{\hspace{0pt}}m{0.115\linewidth}}{Frequency} & TF                                                             & 0.011977                             & 0.007416                             & 0.608795                    & \textbf{0.000011\rlap{\textit{\textsuperscript{**}}}} & \textbf{0.000008\rlap{\textit{\textsuperscript{**}}}} & 0.153479                     \\
                                                              & LF                                                             & 0.095668                             & 0.086258                             & 0.565123                    & \textbf{0.000060\rlap{\textit{\textsuperscript{**}}}} & \textbf{0.000055\rlap{\textit{\textsuperscript{**}}}} & 0.150175                     \\
                                                              & HF                                                             & \textbf{0.003378\rlap{\textsuperscript{*}}}          & \textbf{0.002079\rlap{\textsuperscript{*}}}          & 0.801593                    & \textbf{0.000980\rlap{\textit{\textsuperscript{**}}}} & \textbf{0.000666\rlap{\textit{\textsuperscript{**}}}} & 0.095942                     \\
                                                              & LFn                                                            & 0.216862                             & 0.139111                             & 0.516385                    & \textbf{0.027004\rlap{\textsuperscript{*}}}          & \textbf{0.035141\rlap{\textsuperscript{*}}}          & 0.293047                     \\
                                                              & HFn                                                            & 0.135790                             & 0.087408                             & \textbf{0.032769\rlap{\textsuperscript{*}}} & 0.362291                             & 0.157885                             & \textbf{0.005031\rlap{\textsuperscript{*}}}  \\
                                                              & LF/HF                                                          & 0.135790                             & 0.087408                             & \textbf{0.032769\rlap{\textsuperscript{*}}} & 0.362291                             & 0.157885                             & \textbf{0.005031\rlap{\textsuperscript{*}}}  \\
\hline\hline
\end{tabular}
}
\begin{itemize}[leftmargin=0cm]
    \item[]\emph{**: p-value $<$ 0.001, *: 0.001 $\leq$ p-value $<$ 0.05, NO SYMBOL: p-value $\geq$ 0.05. Bold value indicates significant difference (p-value $<$ 0.05)}
\end{itemize}
\label{table:anova}
\end{table*}


\begin{figure*}[tbp]

 \center

  \includegraphics[width=\textwidth]{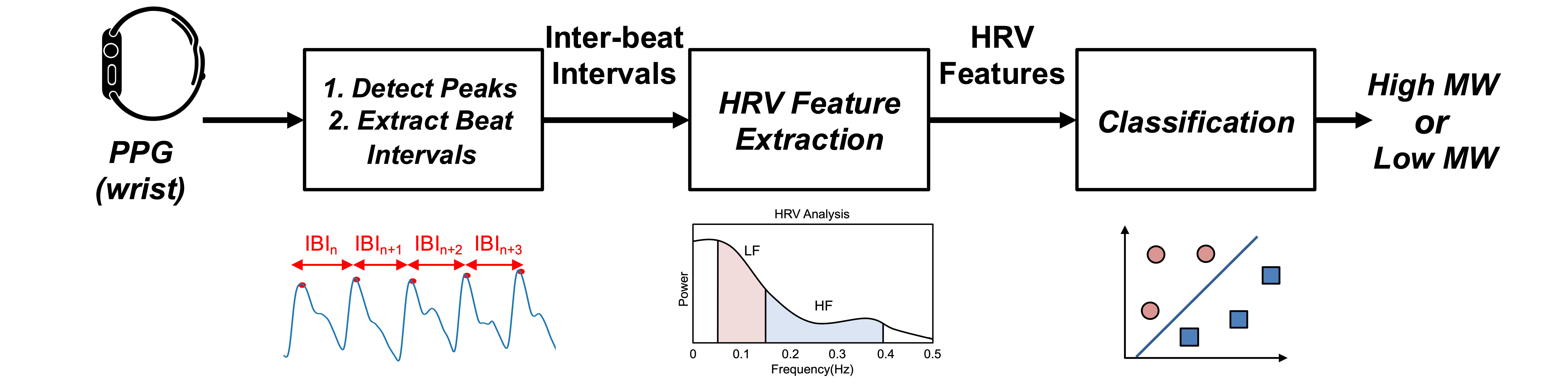}

  \caption{Processing flow of the baseline system.}

  \label{fig:systemflow}

\end{figure*}

\section{Baseline System}
\label{sec:baseline}
In this section, we present a baseline system for predicting the MW(low or high) using physiological signals. Our goal is to provide a reproducible baseline system that can be used to investigate the feasibility of detecting MW using wrist-worn PPG. The signal quality of wrist-worn PPG is usually poor due to motion artifacts. Therefore, our database provides simultaneously recorded ECG and fingertip PPG from Procomp Infinity as references with higher signal quality. With the golden reference, our database offers researchers the opportunity to study enhanced algorithms for MW assessment using wrist-worn PPG and make a fair comparison with others. First, the pre-processing and peak detection methods from the employed signals are described. Then, the extracted features that were used for classification are introduced. Finally, our method and results for single-trial classification using a single modality are presented. The processing flow is shown in Fig.~\ref{fig:systemflow}.

\subsection{Peak Detection}
The following peripheral nervous signals such as ECG, PPG were used to record the subject's MW responses to the stimuli. Several studies~\cite{tao2019systematic} consider that physiological activity is an essential component for MW assessment. For the ECG signal, a bandpass filter was performed between 3 and 45 Hz to enhance the QRS complex peaks. Then an R-peak detection algorithm~\cite{hamilton2002open} was applied, which is an adaption of the popular Pan \& Tompkins algorithm~\cite{pan1985real}. As for the PPG signal, a median filter with the kernel size 13 was applied and followed by a peak detection algorithm ~\cite{peak_detector}. Finally, commonly used features were extracted from each session using 2 minutes windows with 30 seconds overlap. The selection of window size was suggested in ~\cite{castaldo2019ultra} since 2 minutes was the minimum length for various HRV metrics.

\begin{table}[b!]
        \centering
        \caption{Feature list table}
        \begin{tabular}{|c|c|}
        \hline
        \textbf{Feature Type} & \textbf{Features}                                                                                        \\ \hline
        Time-Domain &
          \begin{tabular}[c]{@{}c@{}}SDNN\\ NN50\\ PNN50\\ RMSSD\\ SDSD\\ TINN\\ TriIndex\end{tabular} \\ \hline
        Frequency-Domain            & \begin{tabular}[c]{@{}c@{}}Total Freq. (TF)\\ High Freq. (HF)\\ normalized High Freq. (nHF)\\ Low Freq. (LF) \\ normalized Low Freq. (nLF)\\ LF/HF\end{tabular} \\ \hline
        \end{tabular}
        \label{feature_list}
\end{table}
    
\subsection{HRV Feature Extraction}
 Heart rate variability (HRV) is suggested to correlate with MW changes and tends to diminish when a higher MW is generated~\cite{castaldo2017heart}. HRV is a measure of the heartbeats variance. It is calculated by analyzing the time series of the beat-to-beat intervals from ECG or PPG. There are various methods for extracting HRV, which can basically be divided into time-domain and frequency-domain two analysis methods.   
    
\noindent\textbf{Time-domain Features:}
Several classical metrics have been used as indicators of the total variability of heartbeats using statistical methods~\cite{camm1996heart}, including the standard deviation of inter-beat intervals (IBI) (SDNN), the square root of the mean squared differences between adjacent IBI (RMSSD), the standard deviation of differences between adjacent IBI (SDSD), the count or percentage of successive beats lengths that differed more than 50 ms (NN50, pNN50), the IBI triangular index and the triangular interpolation of IBI histogram (TriIndex, TINN). The extracted time-domain HRV features were listed in Table \ref{feature_list}.

\noindent\textbf{Frequency-domain Features:}
For the frequency-domain features, a spectrum estimation was calculated for the IBI series. We estimated the power spectrum by fast Fourier transform Welch's periodogram techniques. The spectrum was then divided into very low frequency (VLF, 0-0.04 Hz), low frequency (LF, 0.04-0.15 Hz), and high frequency (HF, 0.15-0.4 Hz). Since the signal length for calculating HRV features was 2 minutes, the VLF would be calculated with an incorrect value and therefore not taken. The total power (TF) and LF/HF ratio were also calculated. The LF and HF were also represented as the normalized units (nLF, nHF) to extract the sympatho-vagal component of the HRV better. The extracted frequency-domain HRV features were listed in Table \ref{feature_list}.

\subsection{Classification}
The extracted HRV features would then be used for MW classification. Considering the physiological response was inherently participant independent, the participant-based feature standardization was applied~\cite{MMOD}.  The detection of MW  was a two-class classification problem using the difficulty of the N-back tasks as the ground truth. Since higher $N$ would stimulate a higher MW state, we considered the 2-,3-back tasks as the ``high" MW states and the 0-back task as the ``low" MW state. Support vector machine (SVM)~\cite{SVM} classifiers with a linear kernel were used for MW classification. The adjustable regularization term $C$ of the linear SVM was fine-tuned by grid-search in the range [10e-3, 300]. The parameters of SVM were listed in Table~\ref{table:svm_parameter}. The single modality classifier was trained separately for each channel (ECG, fingertip PPG, and wrist-worn PPG). To validate the classification accuracy in this experiment, we considered two cross-validation methods to train and evaluate classifiers: 

\begin{table}
\centering
\caption{SVM parameters}
\resizebox{\linewidth}{!}{%
\begin{tabular}{>{\hspace{0pt}}m{0.447\linewidth}>{\hspace{0pt}}m{0.44\linewidth}} 
\hline\hline
Parameter                                            & Setups                         \\ 
\hline
Kernel                                               & Linear                         \\
Regularization C & (10e-3 \textasciitilde{} 300)  \\
Class                                                & 2                              \\
Class weight                                         & (0: 0.67, 1: 0.33)             \\
\hline\hline
\end{tabular}
}
\label{table:svm_parameter}
\end{table}

\subsubsection{Leave-one-subject-out (LOSO)~\cite{LOSO}} The LOSO validation was used to ensure that the trained model had no subject bias. In the LOSO validation, the classifier was trained with data from all subjects except one and evaluated on data from excluded subjects. Thus, the training procedure would be finished after all participants' data were evaluated.
\subsubsection{Mixed-Subject 5-fold~\cite{stone1974cross}} The data pool consists of all subjects. We used the 5-fold cross-validation method to split and validate the data pool. The process was repeated 50 times, and the data pool was shuffled to obtain different fold combinations. Finally, the average performance would be presented.

\subsection{Results}
The classification results were shown in Table~\ref{table:baseline result}. The results contained average accuracy and f1-scores (average f1-scores of two classes) for each modality and each validation method, and the standard deviation of each test was also shown after the averaged score. Overall, we observe that the ECG presented the best average performance among the three modalities in both metrics and validation methods. Due to its patch-type measurement, it is least subject to motion interference. The wrist-worn PPG, however, presented the lowest average performance for its poor signal quality. As for the fingertip PPG presented a moderate performance due to its more stable measurement than wrist-worn PPG, and it could even approach the ECG performance in the mixed-subject validation.
Nevertheless, we observed a large gap between wrist-worn PPG and ECG, suggesting that there was still much room for investigating the enhancement of wrist-worn PPG. The ECG could provide a reference for the development of an MW detection system on wearable PPG. While Comparing these two validation methods, the overall performance was similar in terms of average scores. The variance in LOSO was apparently larger than the mixed-subject validation. Since in the LOSO, the left-out subject for testing never appeared in the training data. At the time of testing, unseen data might cause some performance degradation due to differences in individual physiological responses. Therefore, LOSO appeared to a suitable evaluation metric when applying the MW detection system in the real world, as a trained classifier could be used when new subjects needed to be monitored.


\begin{table}[tbp]
\centering
\caption{Evaluation scores of the trained SVM classifier}
\begin{tabular}{clll} 
\hline\hline
Validation                                                                     & \multicolumn{1}{c}{Modality} & \multicolumn{1}{c}{Accuracy (\%)} & \multicolumn{1}{c}{F1-score (\%)}  \\ 
\hline
\multirow{3}{*}{LOSO}                                                          & ECG                          & 71.6 ± 11.1                       & 71.9 ± 10.9                        \\
                                                                               & PPG (fingertip)              & 66.7 ± 11.4                       & 65.0 ± 11.7                        \\
                                                                               & PPG (wrist-worn)             & 59.9 ± 8.8                        & 58.0 ± 9.2                         \\ 
\hline
\multirow{3}{*}{\begin{tabular}[c]{@{}c@{}}Mixed-subject\\5-fold\end{tabular}} & ECG                          & 71.7 ± 5.0                        & 69.0 ± 5.1                         \\
                                                                               & PPG (fingertip)              & 70.8 ± 4.9                        & 69.2 ± 5.0                         \\
                                                                               & PPG (wrist-worn)             & 64.8 ± 5.1                        & 63.2 ± 5.3                         \\
\hline\hline
\label{table:baseline result}
\end{tabular}
\end{table}

\section{Discussion}
\label{sec:discussion}
To our knowledge, MAUS is the first database that includes wristband PPG and fingertips PPG. We provide this database to researchers aiming to develop wearable devices in MW assessment applications. The signal noise, limited quality of subjective rating, and physiological differences between subjects make the assessment of MW challenging. In addition, the gap between ECG and wristband PPG provides an opportunity for further researches to improve wristband PPG performance. The ANOVA analysis of HRV features showed consistent significant trends between ECG and fingertip PPG, while wristband PPG showed differences to these two signals. Considering the complementarity of the significance of features from these signals, this might suggest the possibility of multimodal fusion. We believe our database will enable the researches of developing wristband PPG enhancement algorithms for MW detection.

\section{Conclusion}
\label{sec:conclusion}
This paper presented an open-access database focusing on the study of MW assessment systems for wearable devices. Therefore, a wristband PPG was provided as a representative of wearable devices. In addition, a clinical device that can record ECG, GSR and, fingertip PPG was included in the database as a reference. The N-back tasks were used to stimulate different MW levels. Each participant answered the NASA-TLX questionnaire as subjective workload ratings and the PSQI questionnaire as subjective sleep quality ratings. Besides the experiment procedure description, analysis of subjective rating and physiological signals was also provided.  The database, including raw signals, subjective ratings, task performance, and codes supporting the results of this paper is available at IEEE Dataport \footnote[1]{The Dataset is available at https://ieee-dataport.org/open-access/maus-dataset-mental-workload-assessment-n-back-task-using-wearable-sensor}. Besides, we provide a simple baseline system that could be easily reproduced \footnote[2]{The baseline system code is available at https://github.com/rickwu11/MAUS\_dataset\_baseline\_system}.

\section*{Acknowledgment}
This work was supported in part by the Ministry of Science and Technology of Taiwan under grant MOST-109-2622-8-002-012-TA, and in part by PixArt Imaging Inc., Hsinchu, Taiwan, under grant Pix-108053.

The authors would like to thank PixArt Imaging Inc. that provide PPG watch for data collection. Next, we also grateful to all colleagues and students who contributed to this study, especially to  Hsing-Hao Lee for the help of experiment design from the point view of psychology.
\bibliography{main}
\bibliographystyle{IEEEtran}

\begin{IEEEbiography}[{\includegraphics[width=1in,height=1.25in,clip,keepaspectratio]{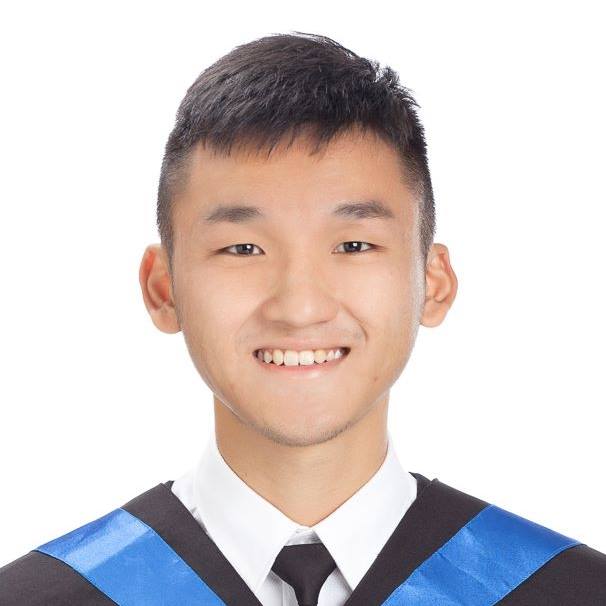}}]{Win-Ken Beh} (IEEE M’19)  received the B.S. degree in electrical engineering from the National Taiwan University,
Taipei, Taiwan, in 2018, where he is currently working toward the Ph.D. degree at the Graduate Institute
of Electronic Engineering, National Taiwan University.
His current research interests include compressive sensing, biomedical signal processing, machine learning, affective computing and VLSI architectures and IC designs.
\end{IEEEbiography}

\begin{IEEEbiography}[{\includegraphics[width=1in,height=1.25in,clip,keepaspectratio]{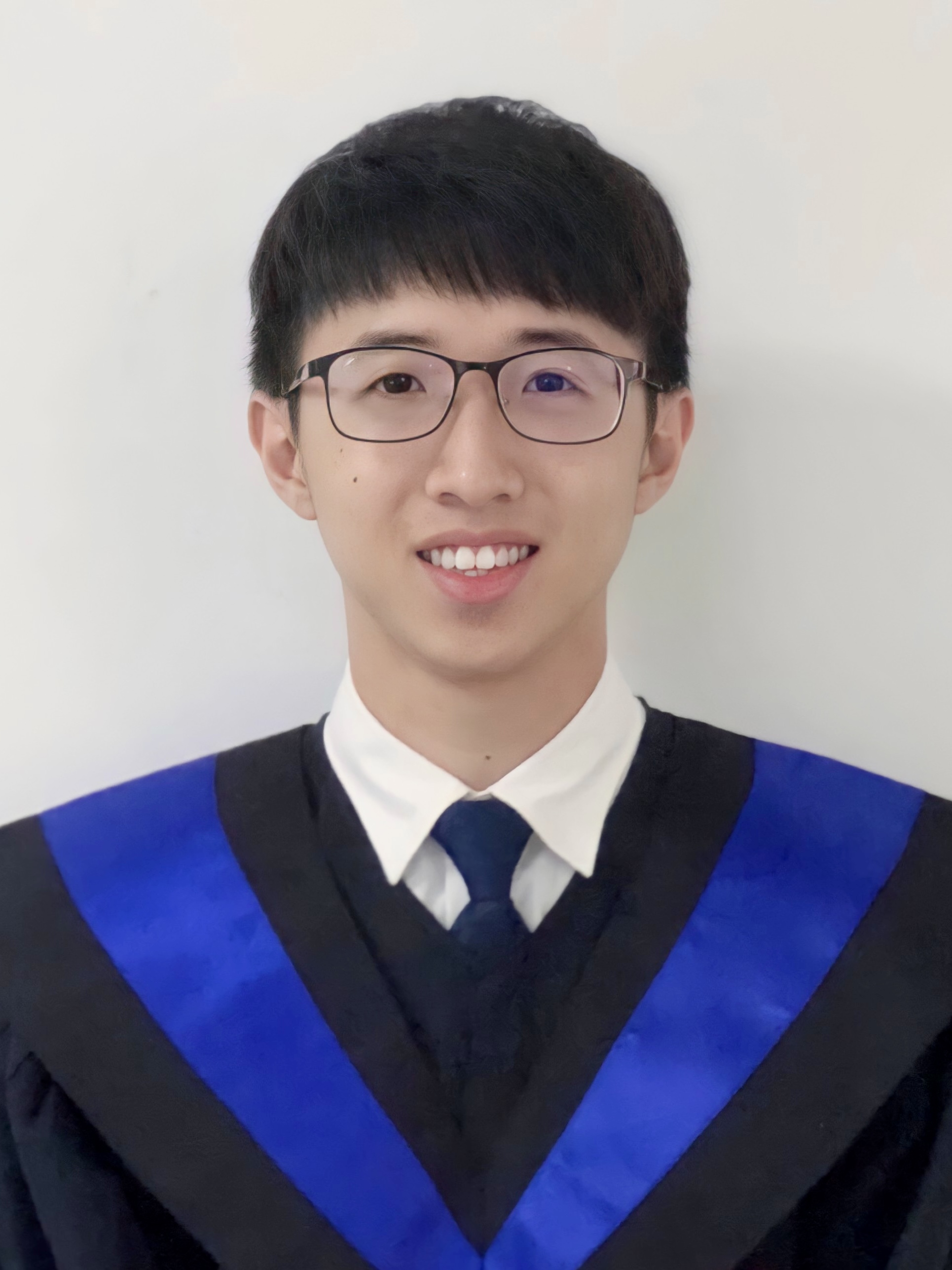}}]{Yi-Hsuan Wu} received the B.S. degree in electrical engineering from the National Taiwan University, Taipei, Taiwan, in 2019, where he is currently working toward the M.S. degree at the Graduate Institute
of Electronic Engineering, National Taiwan University. His current research interests include biomedical signal processing, affective computing and VLSI architectures and IC designs.
 
\end{IEEEbiography}

\begin{IEEEbiography}[{\includegraphics[width=1in,height=1.25in,clip,keepaspectratio]{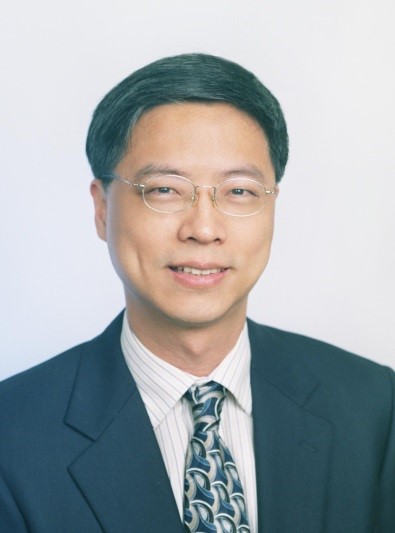}}]{An-Yeu (Andy) Wu } (IEEE M’96-SM’12-F’15)  received the B.S. degree from NTU in 1987, and the M.S. and Ph.D. degrees from the University of Maryland, College Park in 1992 and 1995, respectively, all in Electrical Engineering. 

In August 2000, he joined the faculty of the Department of Electrical Engineering and the Graduate Institute of Electronics Engineering, NTU, where he is currently a Distinguished Professor. His research interests include low-power/high-performance VLSI architectures and IC designs for DSP/communication/AI applications, and adaptive/bio-medical signal processing. He has published more than 260 refereed journal and conference papers in above research areas, together with six book chapters and 24 granted US patents. 

From August 2007 to Dec. 2009, Prof. Wu was on leave from NTU and served as the Deputy General Director of SoC Technology Center (STC), Industrial Technology Research Institute (ITRI), Hsinchu, TAIWAN, supervising WiMAX, Parallel Core Architecture (PAC) VLIW DSP Processor, and Android-based Multicore SoC platform projects. From March 2014 to September 2017, Dr. Wu is in charge of the overall talent cultivation program office in National Program for Intelligent Electronics (NPIE), under sponsorship of Ministry of Education in Taiwan. 

In 2015, Prof. Wu is elevated to IEEE Fellow for his contributions to “DSP algorithms and VLSI designs for communication IC/SoC.” He is elected as a Board of Governor (BoG) Member in IEEE Circuits and Systems Society (CASS) for two terms (2016-2018, 2019-2021). From August 2016 to July 2019, he served as the Director of Graduate Institute of Electronics Engineering (GIEE), National Taiwan University. He recently received 2019 ECE Distinguished Alumni Award from ECE Department of University of Maryland (UMD), and 2019 Outstanding Engineering Professor Award, from the Chinese Institute of Engineers (CIE), Taiwan. He now serves as the Editor-in-Chief (EiC) of IEEE Journal on Emerging and Selected Topics in Circuits and Systems (JETCAS).

\end{IEEEbiography}

\end{document}